\documentclass[fleqn,10pt]{wlscirep}
\usepackage[utf8]{inputenc}
\usepackage[T1]{fontenc}
\usepackage{url,hyperref,lineno,microtype,subcaption,caption}
\usepackage{hyperref}
\usepackage{float}
\usepackage{amsmath,amssymb}
\usepackage{booktabs}
\usepackage{multicol}
\usepackage{xcolor}
\usepackage{booktabs}
\usepackage{multirow}
\usepackage{siunitx}

\title{Predicting survival of glioblastoma from automatic whole-brain and tumor segmentation of MR images}

\author[1,*]{Sveinn Pálsson}
\author[1]{Stefano Cerri}
\author[2]{Hans Skovgaard Poulsen}
\author[2]{Thomas Urup}
\author[3]{Ian Law}
\author[1,4]{Koen Van Leemput}
\affil[1]{Department of Health Technology, Technical University of Denmark, Denmark}
\affil[2]{Department of Oncology, The Finsen Center, Rigshospitalet, Denmark}
\affil[3]{Department of Clinical Physiology, Nuclear Medicine and PET, Center of Diagnostic Investigation, Rigshospitalet, Denmark}
\affil[4]{Athinoula A. Martinos Center for Biomedical Imaging, Massachusetts General Hospital, Harvard Medical School, USA}

\affil[*]{svpa@dtu.dk}

\begin{abstract}

Survival prediction models can potentially be used to guide treatment of glioblastoma patients. 
However, 
currently available MR imaging biomarkers holding prognostic information 
are often challenging to interpret, 
have difficulties generalizing across data acquisitions, or are only applicable to pre-operative MR data. In this paper we aim to address these issues by introducing novel imaging features that can be automatically computed from MR images and fed into machine learning models to predict patient survival. The features we propose have a direct biological interpretation: They measure the deformation caused by the tumor on the surrounding brain structures, comparing the shape of various structures in the patient's brain to their expected shape in healthy individuals. To obtain the required segmentations, we use an automatic method that is contrast-adaptive and robust to missing modalities, making the features generalizable across scanners and imaging protocols.
Since the features we propose do not depend on characteristics of the tumor region itself, they are also applicable to post-operative images, which have been much less studied in the context of survival prediction. Using experiments involving both pre- and post-operative data, we show that the
proposed features carry prognostic value in terms of overall- and progression-free survival, over and above that of conventional non-imaging features.
\end{abstract}

\begin{document}

\flushbottom
\maketitle
%
%
\thispagestyle{empty}

\section*{Introduction}
\label{sec:introduction}

Glioblastoma is the most common malignant primary brain tumor in adults. Prognosis is generally very poor, with a median overall survival (OS) of less than 15 months, and a 5-year OS rate of only 10\%, even when aggressively treated \cite{Louis2007, Gutman2013, Stupp2009, Poulsen2017}. The standard treatment consists of maximal surgical resection followed by radiation therapy and chemotherapy with temozolomide \cite{Stupp2009}. 
Following standard therapy, OS and progression-free survival (PFS) have been shown to correlate with several patient-specific features such as age, performance status and expression of O$^6$-methylguanine-DNA-methyltransferase (MGMT) \cite{Poulsen2017, Michaelsen2013, Hegi2005, Gorlia2008}. However, the prognostic value of these features is still too low to guide treatment choices in individual patients. 

Magnetic resonance (MR) images of glioblastoma patients contain vast amounts of information about the disease, some of which may carry prognostic value. The literature on imaging biomarkers for glioblastoma survival prediction is currently dominated by radiomics \cite{Lambin2012}, an approach in which hundreds or even thousands of features are extracted from delineated tumor regions of MR images, each quantifying some shape, texture, wavelet or histogram property. This approach has shown good performance in predicting survival in many studies \cite{Booth2020, Isensee2017, Weninger2019, Agravat2019, Sun2019, Baid2018, Baid2020, Ingrisch2017}, likely stemming from the correlation between the tumor's texture in MR images and its intratumoral heterogeneity and aggressiveness \cite{BaeR2018, Parekh2019}. However, despite good prediction performance, radiomics suffers from three issues impeding wide-scale practical adoption:
\begin{itemize}
  \item[-]
    \underline{\smash{Lack of interpretability:}}
    Radiomic features, instead of aiming to be interpretable, are designed to be many, to maximize the chance of some having correlation to the target variable. Consequently, many radiomic features are seemingly arbitrary and hard to connect in a meaningful way to the nature of the disease. However,  interpretability of features is important: If a model cannot give biologically meaningful explanations of its predictions, clinicians may not trust the model enough to factor its predictions into their decisions, even if the model is accurate \cite{Shortliffe2018}. Interpretable models may also uncover patterns in the data that give valuable new insight into the disease, and inspire future research. 

  \item[-]
    \underline{\smash{Difficulties generalizing:}}
    The reproducibility of studies using radiomics has been shown to be less than ideal, with results failing to generalize well across scanners and software implementations \cite{Traverso2018,Zwanenburg2020,Welch2019,Gillies2016}. Since many radiomic features depend directly on raw image intensities, they are sensitive to subtle changes in scanning equipment and image acquisition parameters. Furthermore, both textural and shape features depend on the segmentation mask that is used \cite{Orlhac2014}, underlining the importance of using image segmentation methods that are robust with respect to such sources of variation.
 
  \item[-]
    \underline{\smash{Focus on pre-operative data:}}
    Compared to pre-operative images of glioblastoma, relatively little attention has been given to radiomics and other biomarkers in post-operative images. The reason may be that post-operatively, tumor shape and textural features are less easily detectable, as a large part of the tumor is usually removed. Nevertheless, post-operative images are collected closer to the time of disease progression and contain information about the success of operation, making them important to consider in a survival model. While the volume of tumor in post-operative images has been shown to correlate with OS \cite{Ellingson2018,Awad2017}, more advanced imaging biomarkers in post-operative images remain mostly unexplored.
\end{itemize}

In this paper, we propose a method that aims to address these shortcomings. Rather than focusing on in-region radiomic features of the tumor itself, we look at out-of-region features that are more straightforward to interpret and that can readily be applied both to pre- and post-operative data. For this purpose, we take advantage of a recently proposed method to robustly segment dozens of neuroanatomical structures in the presence of tumors~\cite{Agn2019}. Because this method aims to be invariant to imaging variations, it can be directly
applied to data acquired at different centers with different scanners and protocols. 

We demonstrate the resulting surivival prediction method on two fundamentally different datasets: one pre- and one post-operative dataset, each acquired with different scanners, MR contrasts and pre-processing workflows. Our results show that the proposed features improve the performance of survival models for both overall- and progression-free survival, compared to models based only on several previously known prognostic factors. 
To the best of our knowledge, this is the first time a survival model for glioblastoma has been proposed that is 
based on a detailed segmentation of the surrounding brain.

\section*{Methods}
\label{sec:methods}

\begin{figure}[t!]
    \centering
    \includegraphics[width=\textwidth]{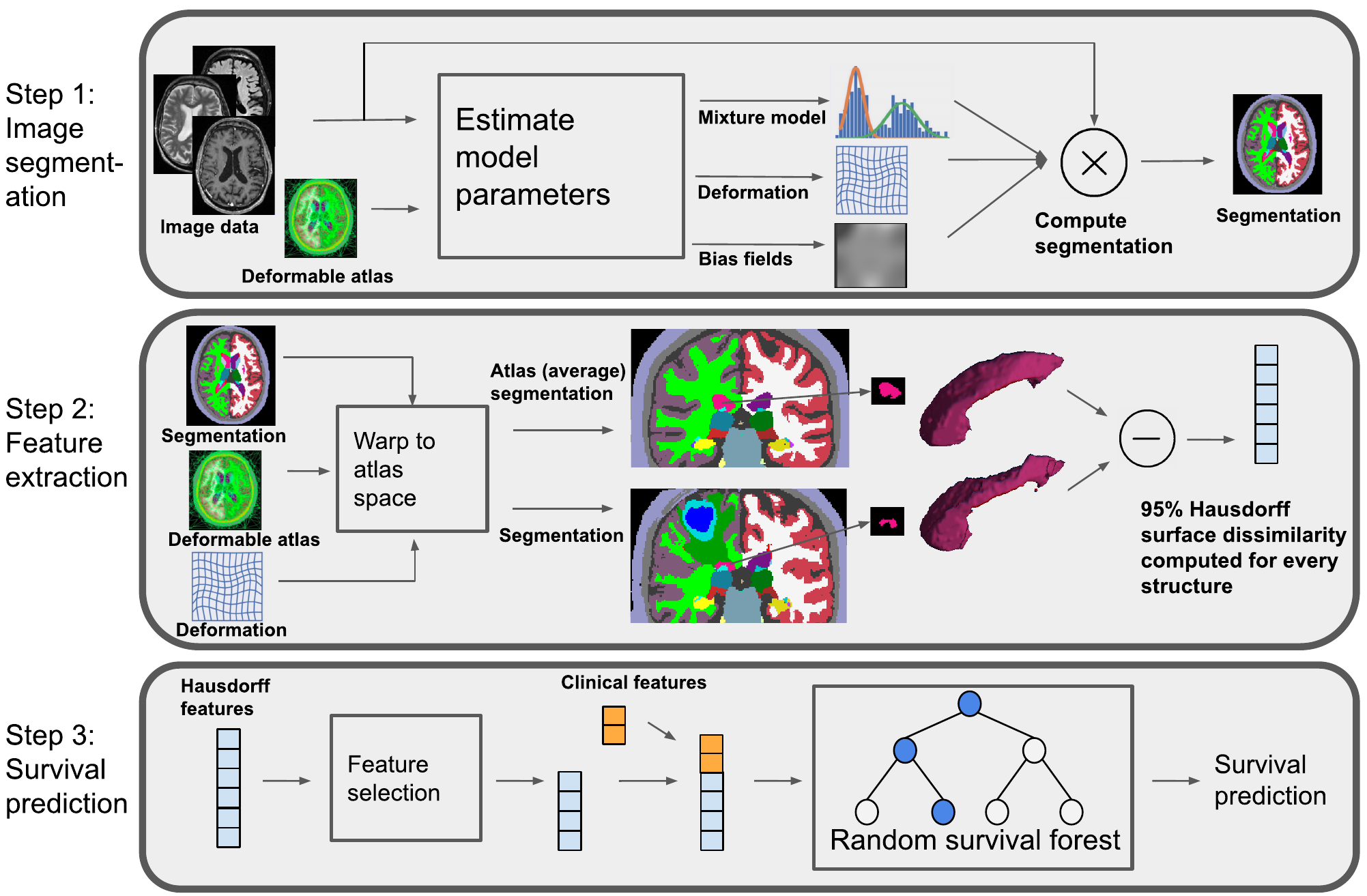}
    \caption{From MR images to survival prediction in three steps: segmentation, feature extraction and survival prediction.
    }
    \label{fig:overview}
\end{figure}

The method we propose for survival prediction consists of three steps, illustrated in Fig.~\ref{fig:overview}. The first step is to segment the images with a contrast-adaptive \textit{whole-brain} segmentation method, simultaneously segmenting dozens of brain structures and the tumor. In the second step, features are computed by comparing each segmented structure to its expected healthy shape using the 95\% Hausdorff distance. In the third step, the extracted features are fed into a feature selector and a survival prediction model.

\subsection*{Image segmentation}
\label{sec:segmentation}

For segmentation we use a method that we recently developed~\cite{Agn2019}, in which three tumor components (edema, enhancing core and non-enhancing core) and dozens of neuroanatomical structures are automatically delineated from a patient's brain MR scan. For post-operative scans, another component is added to capture resection cavity. The method builds on a tool for whole-brain segmentation called Sequence Adaptive Multimodal SEGmentation (SAMSEG), which is distributed with the open-source software suite FreeSurfer~\cite{Freesurfer2012}. It robustly segments head MR scans without any form of preprocessing, using an algorithm that can analyze multimodal data and adapt to variations in contrast due to differences in acquisition hardware or pulse sequences~\cite{Puonti2016}.

SAMSEG is centered on a probabilistic atlas that encodes the spatially varying voxel-wise prior probability of 41 different structures in an average-shaped head~\cite{VanLeemput2008}. This atlas is augmented with a deformation model warping it to match the anatomy of individual subjects, along with models of the MR bias field and of structure-specific intensity profiles. At segmentation time, these models are fitted to the image being segmented, and then used to compute an automatic segmentation (Fig.~\ref{fig:overview}).

For the purpose of segmenting scans with brain tumors, the basic SAMSEG model is further augmented with a spatial regularization model of tumor shape using generative neural networks~\cite{Agn2019}. Although in its original formulation we used convolutional restricted Boltzmann machines~\cite{Lee2011} for this purpose, our current implementation has variational autoencoders~\cite{Kingma2013} since these have a deeper structure and can therefore better represent lesion shape~\cite{Cerri2021}.

\subsection*{Feature extraction}
\label{sec:proposed_features}

Once segmentations are available, we aim to extract features that can sensitively measure the effect the brain tumor has on the shape of the various neuroanatomical structures, compared to those seen in healthy individuals (Fig.~\ref{fig:overview} (Step 2)). To facilitate comparisons between individuals, we compute the features in atlas space, i.e., we warp the automatic segmentations back onto the average-shaped head model by applying the deformation fields that were estimated as part of the segmentation process. The resulting warped, subject-specific segmentations can then be compared to an ``average'' head segmentation that does not take any intensity information into account, obtained by assigning each voxel to the structure with the highest probability in the atlas. We will refer to this ``average'' head segmentation as \emph{the atlas segmentation}. In healthy individuals, the subject-specific warped segmentations will be fairly close to the atlas segmentation in non-cortical structures after warping into atlas space, whereas in brain tumor patients the difference will often be much larger.

\begin{figure}[H]
    \centering
    \includegraphics[width=\textwidth]{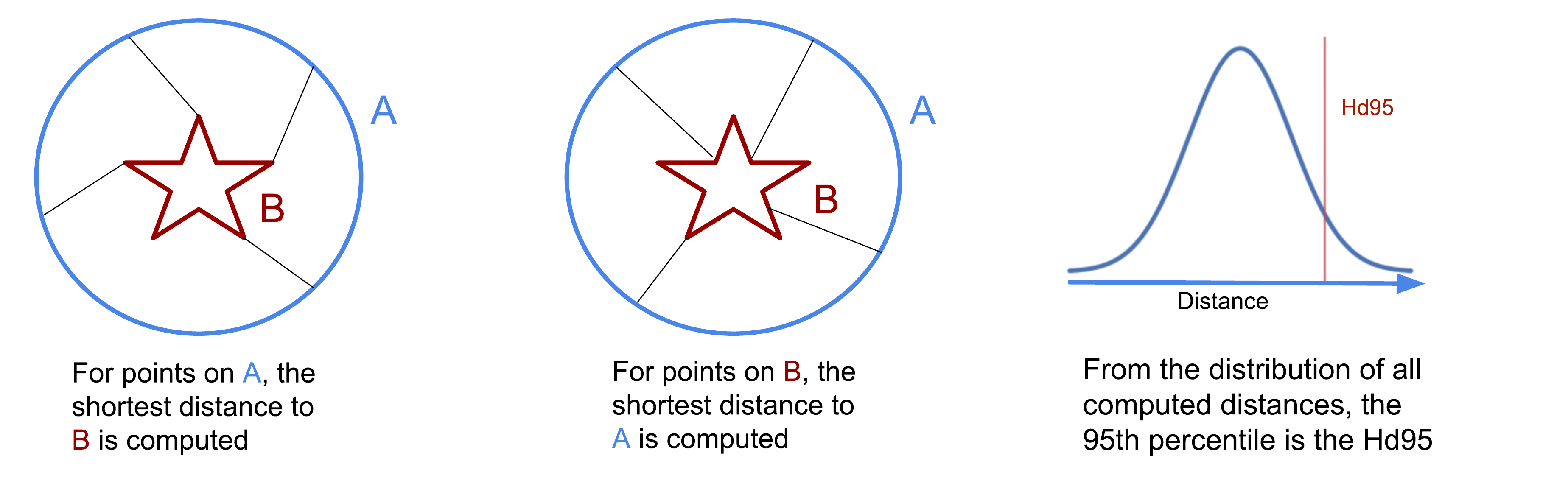}
    \caption{An illustration of how the Hausdorff 95\% distance (Hd95) is computed between two example shapes.}
    \label{fig:hd95}
\end{figure}

In order to quantitatively compare the two segmentations, we compute a robust version of the Hausdorff distance~\cite{Huttenlocher1993} for each of 26 relevant structures: Accumbens area (L\&R), amygdala  (L\&R), brain stem, caudate (L\&R), cerebellum cortex (L\&R) , cerebral cortex (L\&R), hippocampus (L\&R), lateral ventricle (L\&R), optic chiasm, pallidum (L\&R), putamen (L\&R), thalamus (L\&R), ventral diencephalon (L\&R), 3rd- and 4th-ventricles. The Hausdorff distance measures the distance between the outer borders of a pair of segmentation masks; its robust version is an often-used metric to quantify the performance of automatic segmentation methods with respect to manual ``ground truth'' delineations performed by human experts~\cite{Menze2014}. Let $A$ and $B$ denote the outer border of the segmentation masks of a particular brain structure, obtained from the atlas and warped segmentation, respectively. The Hausdorff distance computes, for all voxels on the border $A$, the shortest Euclidean distance to voxels on the border $B$, and vice versa, and returns the maximum value over all the computed distances. Because the maximum distance is highly sensitive to outliers, the robust version instead returns the 95th percentile of the distances (Fig.~\ref{fig:hd95}). The robust version is often called the 95\% Hausdorff distance but for short, will be referred to as Hd95 throughout the rest of the paper.

In cases where no voxel is assigned to a structure when obtaining the automatic segmentation, the Hd95 is not defined. In such cases, we instead use a single voxel located at the center of mass of the corresponding atlas segmentation.

\begin{figure}[H]
    \centering
    \includegraphics[width=\textwidth]{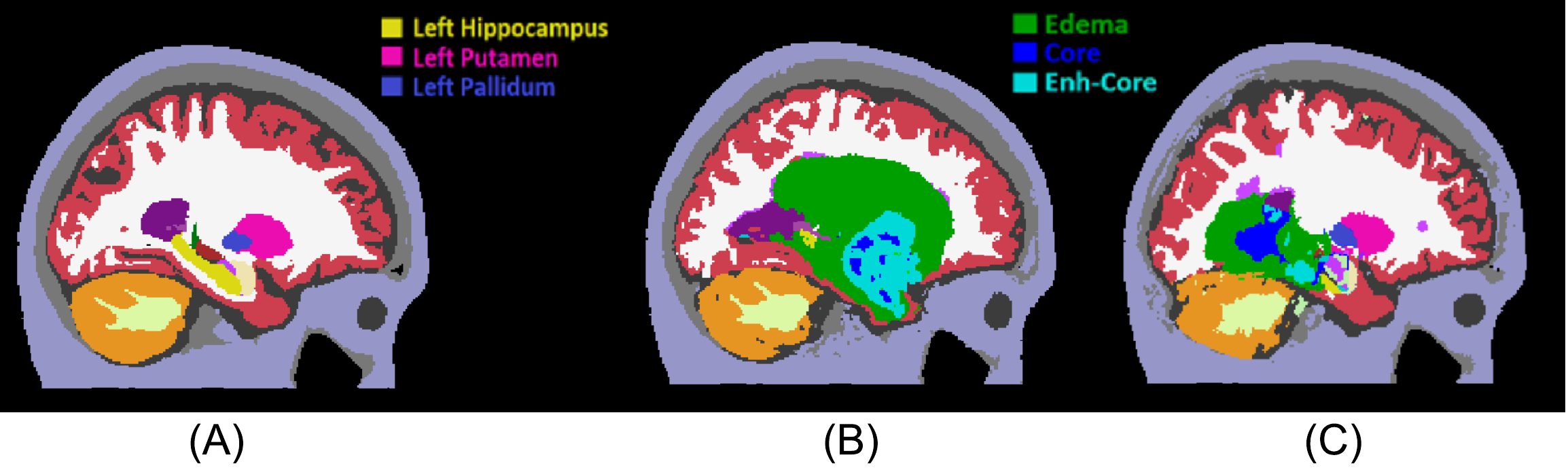}
    \caption{The atlas segmentation reflecting average anatomy (A) and 
    automatic
    brain segmentations of two subjects with glioblastoma (B-C). The subject in (B) has highly deformed left hippocampus, pallidum and putamen, which is reflected in high Hd95 values for these structures. 
    While the hippocampus in (C) is also deformed, the pallidum and putamen are largely unaffected. }
    \label{fig:hd95example}
\end{figure}

For an example of how Hd95 captures the deformation of brain structures, Fig.~\ref{fig:hd95example} shows two subjects with glioblastoma (Fig. ~\ref{fig:hd95example} (B-C)) and the corresponding atlas segmentation (Fig.~\ref{fig:hd95example} (A)) for comparison. The tumor in figure~\ref{fig:hd95example} (B) has a clear effect on the shape of the left hippocampus, putamen and pallidum, with an estimated Hd95 of 21.7, 28.5 and 49.3 [mm], respectively. While also showing a clear deformation of the left hippocampus, the left pallidum and putamen in Fig.~\ref{fig:hd95example} (C) seems largely unaffected, with Hd95 of 24.6, 2.5 and 2.2 [mm], respectively. 

The proposed Hd95 features contain some information about where the tumor is located in the brain and its size, both of which have been studied before and shown to carry prognostic value~\cite{Awad2017, Gorlia2012, Chaichana2008, Gorlia2008, Abou2019, Yersal2017,Poulsen2017}. To verify that any prognostic value of our features is not solely based on tumor size and location, in our experiments we also evaluate the performance of our survival prediction models when they are trained directly on the estimated tumor size and the center-of-mass (CoM) coordinates of the whole tumor (defined as the set of voxels assigned to any tumor component). The contrast-enhancing tumor volume (CEV) is the tumor size definition most widely used clinically, but we will also consider the volume of each tumor component (TCV), including resection cavity in case of post-operative images.

\subsection*{Survival prediction}
\label{sec:survival_prediction}

Survival predictions models were trained following a standard machine learning workflow. The workflow consists of feature selection to remove uninformative features, and subsequent fitting of a survival prediction model to the remaining features (see Fig.~\ref{fig:overview} (Step 3)).

For feature selection, we used the univariate Cox proportional hazards (Cox PH) model \cite{Cox1972} (implemented in python~\cite{lifelines}), considering one feature at a time and retaining it if its coefficient is significantly nonzero. 
We used two sided P-values and considered $P < 0.05$ statistically significant.

A random survival forest (RSF)~\cite{Ishwaran2008} was used as the prediction model (implemented in python~\cite{sksurv}). RSF extends the random forest model~\cite{Breiman2001} to handle right-censored data, i.e., subjects who had not yet died by the end of the study -- knowing that these subjects survived at least until their recorded time still contributes to fitting the RSF parameters. The RSF is an ensemble of trees whose leaf nodes estimate the subject's survival function from training data seen by the node. The survival prediction for a subject is taken as the expected survival of the average survival function across all leaf nodes the subject visits. Due to the small number of subjects in our datasets, we did not optimize over the RSF hyperparameters but left them at the default setting in the survival analysis software: 100 trees, no maximum depth, 6 subjects minimum to split a node and minimum 3 subjects in leaf nodes. Models were trained via K-fold cross-validation where K was chosen such that in each fold, 5 subjects were left out while the model was trained on the remaining N-5 subjects (K$=N/5$); the model was then used to predict survival of the 5 left-out subjects. We repeated this procedure 100 times for more accurate estimation of model performance.

\section*{Experiments and Results} \label{sec:experiments_results}

To demonstrate the versatility and reproducibility of the proposed method across data acquisitions, we performed experiments on two fundamentally different datasets: an in-house dataset of post-operative scans, and a  publicly accessible dataset of pre-operative scans. Here we first describe these datasets, and subsequently present results for each.

\subsection*{Datasets}
\label{sec:study_population}

\subsubsection*{Copenhagen dataset (post-operative)}
Our primary focus is on a set of post-operative scans acquired at Rigshospitalet, Copenhagen. It contains MR scans of 146 histologically verified glioblastoma patients, diagnosed in the period September 2011 - April 2014. Permission for data collection was given from the Danish Data Protection Agency (2006-41-6979). Each patient received radiation therapy with concomitant and adjuvant temozolomide (see~~\cite{Poulsen2017} for details about treatment). OS and PFS were recorded in months for all subjects with 14 and 6 censored subjects (i.e. still alive/non-progression at the end of the study), respectively.

\begin{figure}[t!]
    \centering
    \includegraphics[width=0.9\textwidth]{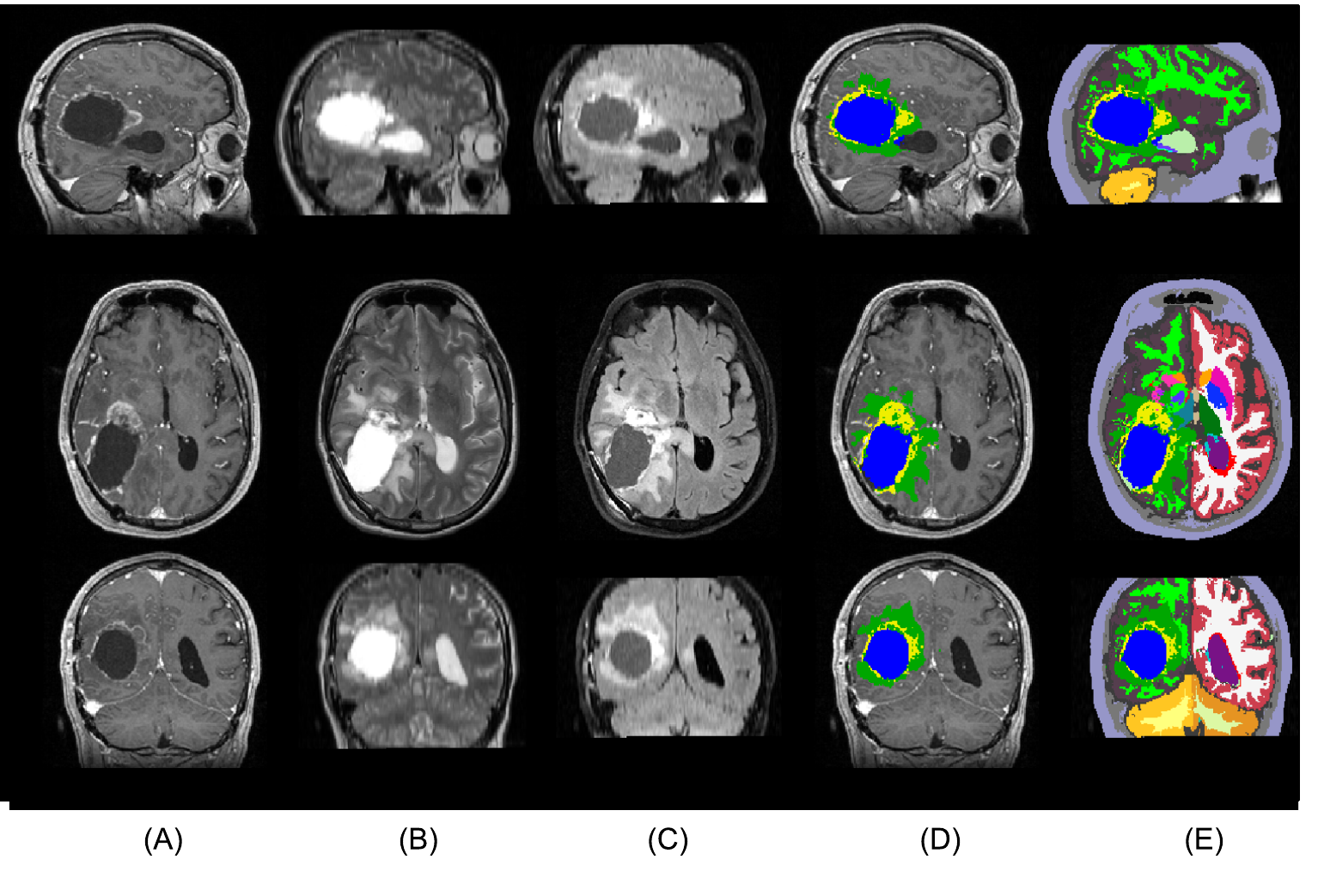}
    \caption{A sample from the Copenhagen (post-operative) dataset. From top to bottom: sagittal, axial and coronal view. The columns show (A) T1c, (B) T2, (C) FLAIR, and (D-E) the automatic segmentation output. (D) shows the tumor components only, while (E) shows the full segmentation output. The tumor components in (D-E) are edema (green), enhancing core (yellow) and non-enhancing core (blue). Resection cavity is shown in light green color in the sagittal view of (E).
    }
    \label{fig:rh_mri}
\end{figure}

MR scans were acquired for radiation planning 2-3 weeks post-operatively. The acquired MR modalities included 3D T1 (MPRAGE) post-administration of gadolinium contrast (T1c), T2 and FLAIR (Fig.~\ref{fig:rh_mri} (A-C)), using a 1.5T Siemens Espree scanner. The T1c scans were acquired using a voxel size of $0.5 \times 0.5 \times 1.0 \text{~mm}^3$ (matrix size $384 \times 512 \times 176$); the FLAIR scans with a voxel size of $0.45 \times 0.45 \times 3.3 \text{~mm}^3$ (matrix size $448 \times 512 \times 40$); and the T2 scans using a voxel size of $0.3 \times 0.3 \times 3.3 \text{~mm}^3$ (matrix size $672 \times 768 \times 39$). As the only form of pre-processing, intra-subject registration and resampling to 1mm$^3$ resolution was performed using FLIRT \cite{Jenkinson2002}. Three out of the 146 subjects were excluded as their post-operative MR data was unavailable. Out of the remaining 143 subjects, 11 were missing FLAIR scans and 3 were missing T2. However, our segmentation algorithm is robust with respect to missing modalities, allowing all 143 subjects to be included in the study. 

Additional features recorded in the clinic were the patient's age, performance status and MGMT protein status. As mentioned in the introduction, these are features that have been previously shown to have prognostic value and are thus commonly considered for radiotherapy planning. We will refer to this set of variables as the ``clinical features''. 

\begin{figure}
    \centering
    \includegraphics[width=\textwidth]{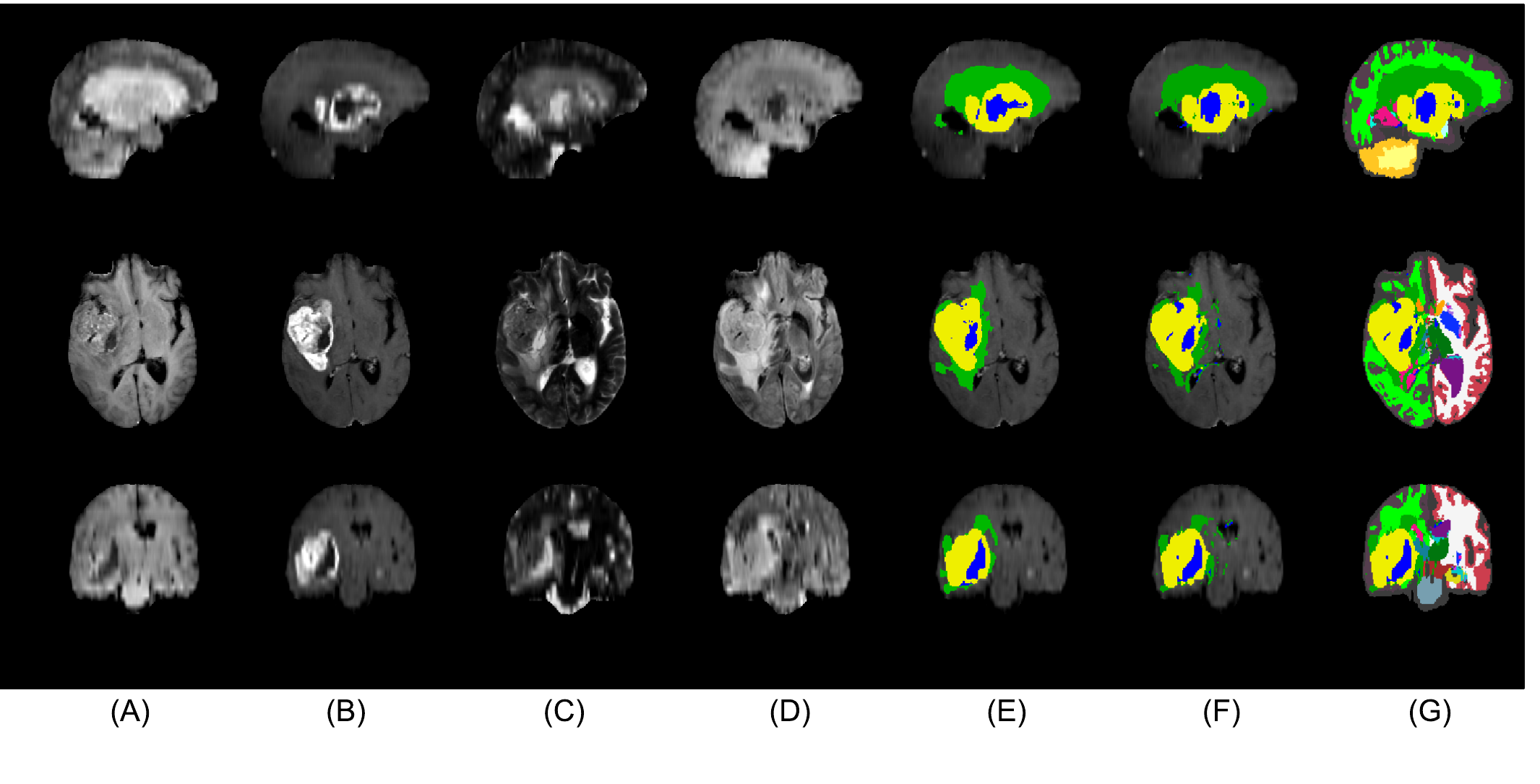}
    \caption{A sample from the BraTS20 (pre-operative) dataset. From top to bottom: sagittal, axial and coronal view. The columns show (A) T1, (B) T1c, (C) T2, (D) FLAIR, (E) manual segmentation of tumor, and (F-G) the automatic segmentation output. (F) shows the tumor components only, while (G) shows the full segmentation output. Some major differences to the Copenhagen (post-operative) dataset (see Fig.~\ref{fig:rh_mri}) can be seen in this figure.
    }
    \label{fig:brats_mri}
\end{figure}

\subsubsection*{BraTS20 dataset (pre-operative)}

To test the reproducibility of the methods we propose, we also applied them to a fundamentally different (namely, pre-operative) dataset, obtained with other acquisition settings and preprocessing steps. The Multi-modal Brain Tumor Segmentation Challenge 2020 (BraTS20) \cite{Bakas2018, Bakas2017, Bakas2017_2, Bakas2017_3} released a publicly available set of 235 high grade glioma subjects with overall-survival times. This dataset contains both glioblastoma and anaplastic astrocytoma \cite{Menze2014}, although more detailed information on the subjects' sub-classification is not provided. For each subject, information on their age and OS is provided, but PFS or other clinical features are not available. None of the 235 subjects are censored.

The MR scans originate from multiple clinics and were acquired on different scanners, with magnetic field strengths of 1.5T and 3T. For each subject, the dataset contains a T1 pre- and post-administration of gadolinium contrast, a T2 and a T2 FLAIR scan (Fig.~\ref{fig:brats_mri} (A-D)). In a pre-processing step, the images were aligned to a brain template, interpolated to 1mm$^3$ isotropic resolution and skull-stripped by the challenge organizers~\cite{Menze2014,Bakas2018}. Despite differences in available MR contrasts and in pre-processing compared to the Copenhagen dataset, our segmentation method did not need adjustment to handle the BraTS20 data. 

\subsection*{Results on the Copenhagen dataset}
\label{sec:results_cph}

We present several different aspects of the proposed prediction method. First, we look at which Hd95 features were automatically selected for inclusion in the survival models. We then make a comparison between models trained on different feature sets, and we test the proposed method's ability to stratify patients into high- and low-risk groups based on their predictions. Finally, we evaluate the discriminative power of individual features for predicting short and long survival.

\subsubsection*{Feature selection}
\label{sec:featureSelectionResultsCPH}

Rather than reporting on Hd95 features selected within each fold during cross-validation, for conciseness here we present results of the Cox PH feature selection method on the entire cohort. Although this introduces information leakage between training and test sets, in our experiments we found that the selected features across folds were highly consistent, thus having minimal impact on the overall prediction performance (details provided in Appendix A). As shown in Table~\ref{tab:selected_features}, our feature selection resulted in 10 retained Hd95 features for OS prediction, and 4 for PFS. 

On the Copenhagen dataest, our feature selection resulted in 10 retained Hd95 features for OS prediction, and 4 for PFS, while 4 were retained for OS prediction on the BraTS20 dataset.

\begin{table}[H]
    \begin{tabular}{llll}
    \toprule
    Feature &  OS (Copenhagen) &  PFS (Copenhagen) & OS (BraTS20) \\
    \midrule
Amygdala 				& \checkmark (L)   &    & \checkmark (L) \\
3rd-Ventricle 			& \checkmark 	   & 	& \\
Hippocampus             & & & \checkmark (L) \\
Lateral ventricle 		& \checkmark (L)   &    \\
Pallidum 				& \checkmark (L,R) & \checkmark (L)  & \checkmark (L) \\
Putamen 				& \checkmark (L)   & \checkmark (L)  & \checkmark (L) \\
Thalamus 				& \checkmark (L,R) & \checkmark (L) & \\
Ventral diencephalon 	& \checkmark (L,R) & \checkmark (L) & \\

    \end{tabular}
    \caption{Brain structures whose Hd95 feature was selected by the feature selection method are marked with a check mark, accompanied by L and R denoting left and right sided structures. Shown for both the Copenhagen and BraTS20 datasets.
    }
    \label{tab:selected_features}
\end{table}

\subsubsection*{Subject-level prediction performance}

To evaluate the prognostic value of the Hd95 features, in this section we investigate the performance of RSF prediction models trained on different sets of input features. In particular, we are interested in the comparison of models trained with the clinical features alone; the Hd95 features alone; and the combination of both. In addition, we compare with models that include tumor size (either TCV or CEV) and center-of-mass (CoM) as input features, as well as with models that only use age as the clinical variable. Note that feature selection was only performed on the Hd95 features as the clinical, size and location features have all been previously shown to carry prognostic value~\cite{Awad2017, Gorlia2012, Chaichana2008, Gorlia2008, Abou2019, Yersal2017,Poulsen2017}.

The first two columns of Table~\ref{tab:results} show performance of the RSF models, computed from the cross-validated predictions on the Copenhagen dataset. To quantify the performance of any given model, its predictions were compared with the ground truth survival times using Harrell's concordance index (C-index) \cite{Harrell1982}. The C-index computes the probability that for a pair of randomly selected subjects, their predicted survival is correctly ordered with respect to their true survival times. A C-index value of 1 means perfect prediction performance while 0.5 is the expected result of blindly guessing. The reported C-index is the average over the 100 repetitions of cross-validation, accompanied by the 95\% confidence interval of the mean in brackets.

The best model for OS was achieved by combining the proposed features with the previously known prognostic clinical features: further addition of CoM, TCV and CEV did not provide significant improvement. Individually, the clinical, size and location features all showed lower performance than the Hd95 features for OS prediction, and when combined they achieved only 0.624 C-index, compared to the 0.670 C-index when Hd95 was also included. The Hd95 features thus seem to bring prognostic value that is not contained in simple size and location based features. 

For PFS, the best model was achieved by combination of Hd95, clinical, CoM and CEV, achieving a C-index of 0.637. Individually, the CoM was the best predictor of PFS and combining it with clinical and size features achieved a C-index of 0.622. The benefit of including the Hd95 features is clear for PFS, but is considerably lower than for OS.

\begin{table}[H]
    \begin{tabular}{lccc}
    \toprule
    \multirow{2}{*}{Features} &
    \multicolumn{2}{c}{Copenhagen} &
    \multicolumn{1}{c}{BraTS20} \\
     &   OS &  PFS &  OS  \\
    \midrule
    
Hd95 + Clinical + CoM + CEV						& \textbf{0.670} (0.668 - 0.671) & \textbf{0.637} (0.634 - 0.641) & 0.619 (0.618 - 0.620) \\
Hd95 + Clinical + CoM + TCV				     	&  0.657 (0.655 - 0.659) & 0.629 (0.624 - 0.634) & 0.612 (0.611 - 0.614)\\
Hd95 + Clinical	+ CEV							& 0.666 (0.665 - 0.668) & 0.597 (0.595 - 0.600) & \textbf{0.631} (0.630 - 0.632) \\
Hd95 + Clinical     							&  \textbf{0.669} (0.668 - 0.671) & 0.614 (0.612 - 0.616) & 0.612 (0.611 - 0.613)\\
Hd95 + CoM + CEV								& 0.635 (0.633 - 0.637) & 0.629 (0.625 - 0.634) & 0.564 (0.562 - 0.565) \\
Hd95 + CEV										& 0.643 (0.641 - 0.644) & 0.580 (0.575 - 0.584) & 0.594 (0.593 - 0.595) \\
Clinical + CoM + CEV 							& 0.624 (0.623 - 0.625) & 0.622 (0.618 - 0.626) & 0.599 (0.598 - 0.600) \\
Clinical + CoM + TCV							&  0.595 (0.592 - 0.598) & 0.613 (0.608 - 0.618) & 0.600 (0.598 - 0.602)\\
Clinical + CEV		 							& 0.591 (0.589 - 0.593) & 0.567 (0.563 - 0.572) & 0.616 (0.614 - 0.617) \\
CoM + CEV			 							& 0.548 (0.545 - 0.551) & 0.605 (0.599 - 0.611) & 0.517 (0.516 - 0.518) \\
CoM + TCV 										&  0.540 (0.537 - 0.543) & 0.617 (0.612 - 0.622) & 0.536 (0.533 - 0.539)\\
Hd95     										&  0.644 (0.643 - 0.646) & 0.552 (0.548 - 0.556) & 0.571 (0.570 - 0.572)\\
Clinical 										&  0.574 (0.572 - 0.576) & 0.524 (0.522 - 0.527) & 0.581 (0.579 - 0.583)\\
CoM 					 						& 0.550 (0.548 - 0.551) & 0.591 (0.588 - 0.594) & 0.504 (0.503 - 0.506) \\
CEV 					 						& 0.479 (0.476 - 0.482) & 0.551 (0.547 - 0.554) & 0.534 (0.533 - 0.535) \\
TCV    			     							& 0.525 (0.522 - 0.528) & 0.574 (0.570 - 0.577) & 0.553 (0.551 - 0.555) \\
Age      										&  0.509 (0.506 - 0.513) & 0.519 (0.516 - 0.522) & 0.581 (0.579 - 0.583)\\

    \end{tabular}
    \caption{
    Prediction performance measured with the C-index for models trained on several different sets of features. The first two columns show results for OS and PFS prediction on the Copenhagen post-operative dataset, whereas the last column contains OS prediction performance on the BraTS20 pre-operative dataset. Note that the clinical features for the Copenhagen dataset include age, performance status and MGMT methylation, while the available clinical features for the BraTS dataset consist only of age. Including the proposed Hausdorff (Hd95) features in the survival model provides an improvement in prediction performance over models that only consider conventional clinical features, tumor size and location. 
    }
    \label{tab:results}
\end{table}

\subsubsection*{Risk group stratification}
\label{sec:groupAnalysis}

\begin{figure}
  \includegraphics[width=\linewidth]{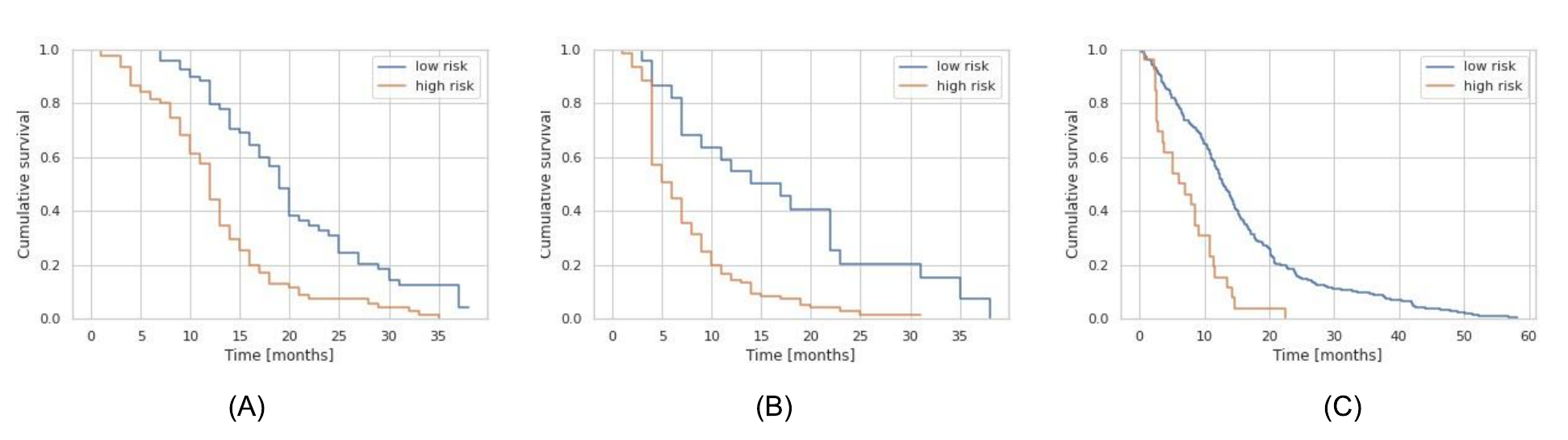}
  \caption{Kaplan-Meier curves showing the cumulative survival (fraction of the population alive/without progression at a given time),
  for (A) OS and (B) PFS in the Copenhagen dataset and (C) for OS in the BraTS20 dataset. The two survival groups in each figure are obtained by splitting the cohort based on the predicted survival at the threshold that best separates the cohort.
  }
  \label{fig:survival_groups}
\end{figure}

Here we demonstrate that the proposed survival models can be used to stratify patients into low- and high-risk groups. For this purpose, a threshold was selected by searching, among the predictions for all Copenhagen patients, for the value that best separates the dataset in terms of the recorded survival \cite{Contal1999, BaeR2018}. Separation quality was measured with the log-rank test \cite{Mantel1966}, which tests the hypothesis that two groups have the same survival distribution. The prediction value yielding the lowest P-value (of the log-rank test) was chosen as the threshold separating the low- from the high-risk patients. Visualization of the resulting groups, using the RSF models trained on the combination of clinical and selected Hd95 features as prediction models, is shown with Kaplan-Meier survival curves \cite{kaplan1958} for OS and PFS in Fig.~\ref{fig:survival_groups} (A) and (B), respectively. 

We further computed the corresponding hazard ratio for the obtained splits (ratio of hazard rates between the two groups under the proportional hazards assumption \cite{Sashegyi2017}), using the univariate Cox proportional hazards model where the input covariate was the group membership. In addition to the hazard ratio, its 95\% confidence interval and log-rank P-value were also computed. For OS, the hazard ratio was $2.65$ ($1.85 - 3.79$), $P=10^{-8}$  and for PFS the hazard ratio was $1.85$ ($1.7-4.78$), $P=10^{-5}$. These results show that our survival models can stratify patients into significantly different survival groups for both OS and PFS.

\subsubsection*{Prognostic potential of individual features}
\label{sec:featureImportance}

The Hd95 features we propose have a clear biological interpretation: higher values reflect more severe deformation in the corresponding brain structures. To test the intuition that highly deformed individual structures are associated with poor outcomes, we concentrated on subjects with very high deformations and tested to what degree their survival differs from that of the remaining subjects. Specifically, for each of the 26 brain structures for which we computed Hd95 features, we split the subjects into two groups according to whether or not they are in the highest 10\% range of feature values. We then computed 1. the percentage of short survivors (below the median survival of the cohort) among the subjects in the highest 10\% range, and 2. the log-rank test between the two groups.

The results of this experiment are in listed in Table~\ref{tab:survival_percentage} for structures where the log-rank P-value was significant. The results show that for several brain structures, high Hd95 value is a strong predictor of short survival. The best predictor of OS was deformation of the left lateral ventricle, where 92\% of the subjects with the most deformation were short survivors. For PFS, the best predictor was the deformation of the left thalamus, with 91\% of the subjects with the most deformation of that structure being short survivors.

\begin{table}[H]
    \centering
    \begin{tabular}{lcccccc}
    \toprule
    \multirow{3}{*}{Hd95 features} &
    \multicolumn{2}{c}{OS (Copenhagen)} &
    \multicolumn{2}{c}{PFS (Copenhagen)} &
    \multicolumn{2}{c}{OS (BraTS20)} \\
    & \% short & $P$   &  \% short & $P$  &  \% short & $P$ \\
    \midrule
    Left lateral ventricle & 92  &  $1\times10^{-3}$  & 71  &  $4\times10^{-2}$  & -  & - \\
Left putamen	     	& 85  &  $2\times10^{-3}$  & 71  &  $2\times10^{-3}$  & 71 &  $2\times10^{-3}$  \\
Left pallidum 		 	& 83  &  $4\times10^{-3}$  & 71  &  $7\times10^{-3}$  & 67 &  $5\times10^{-3}$  \\
Left thalamus 	     	& 82  &  $7\times10^{-3}$  & 91  &  $2\times10^{-2}$  & 65 &  $2\times10^{-2}$ \\
Left ventral diencephalon
  	 	& 77  &  $6\times10^{-3}$  & 69	 & 	$2\times10^{-2}$  & 81 &  $2\times10^{-4}$   \\
4th ventricle 		    & 58  &  $4\times10^{-2}$  & -	 &  -				  & -  &  - \\
Left amygdala 		 	& -   &  - 			       & -   &  -   			  & 79 &  $1\times10^{-2}$  \\
Left hippocampus   	& -   &  - 			       & -   &  -   			  & 75 &  $7\times10^{-4}$  \\

    \end{tabular}
    \caption{Percentage of short survivors
    among the subjects in the highest 10\% range of individual Hd95 feature values. The table also shows the P-value of a log-rank test between the survival times of subjects within and outside the highest 10\% range. 
    Brain structures where the log-rank P-value $>0.05$ are omitted.
    }
    \label{tab:survival_percentage}
\end{table}

\subsection*{Results on BraTS20 dataset}
\label{sec:results_brats}

The same methods were applied to the BraTS20 (pre-operative) dataset, for which our goal was to predict the OS only. 

\subsubsection*{Feature selection}
\label{sec:featureSelectionResultsBraTS}
Feature selection on the full BraTS20 cohort resulted in selection of 4 Hd95 features: left putamen, left pallidum, left hippocampus and the left amygdala (listed in Table~\ref{tab:selected_features} for comparison with Copenhagen dataset). Similarly to the Copenhagen dataset results, selecting features within each fold of cross-validation resulted in mostly the same features being chosen (see details in Appendix A). Note that three of the selected features on the BraTS20 data were also selected for OS prediction on the Copenhagen dataset 
(vs.~two for PFS, cf.~Table~\ref{tab:selected_features}).

\subsubsection*{Subject-level prediction performance}
\label{sec:modelComparisonBraTS}
Model comparison to test whether the Hd95 features contain prognostic information not included in the clinical, size or location data was done in the same manner as with the Copenhagen dataset. An important difference is that the only clinical data available here is the subject's age, while the Copenhagen data also included MGMT and performance status. 

As shown in the last column of Table~\ref{tab:results}, the best OS prediction model was obtained with a combination of Hd95, CEV and age. This is largely in line with the results we obtained for OS prediction on the Copenhagen data (cf.~first column of Table~\ref{tab:results}), where the best models were the ones combining Hd95 with other features. The results for size and location features are similar between the datasets: neither are good OS predictors individually. However, individually, here the age was the best feature, achieving a C-index of 0.581, which is substantially higher than in the Copenhagen dataset where age alone only achieved 0.509. Although the performance of the proposed Hd95 features and CEV individually was quite low (0.571 and 0.534, respectively), combining them both with the age achieved the best C-index of 0.631. While one of the best models for OS prediction on the Copenhagen dataset was the model combining Hd95 with clinical features, that specific combination only achieved a C-index of 0.612 on the BraTS20 dataset. Nevertheless, this is still an improvement over considering either of the two feature sets individually. It is further worth noting that the age is the only clinical feature provided in the BraTS20 dataset -- addition of MGMT and performance status information could improve the performance and possibly outperform the model using Hd95, CEV and age also here.

\subsubsection*{Risk group stratification}
\label{sec:groupAnalysisBraTS}

As in the Copenhagen dataset, the prediction model trained on the combination of clinical and selected Hd95 features was used to stratify the BRaTS20 cohort. The Kaplan-Meier curves in Fig.~\ref{fig:survival_groups} (C) show the proportion of subjects alive at any given time point for the two resulting groups. The corresponding hazard ratio was $2.81$ ($1.84 - 4.29$) and log-rank P-value $5\times 10^{-7}$, indicating that the two resulting groups have significantly different OS.

\subsubsection*{Prognostic potential of individual features}
\label{sec:featureImportanceBraTS}
To explore to what degree individual Hd95 features can predict short survival in the BraTS20 dataset, we repeated the experiment of exploring the percentage of short survivors among the subjects with the most deformed brain structures. As shown in Table~\ref{tab:survival_percentage}, the highest 10\% range of feature values is predictive of short survival for several structures. Compared to our results on the Copenhagen dataset, two new structures show high predictive power in the BraTS20 data. 

\section*{Discussion}
\label{sec:discussion}

In this paper we have proposed a new set of imaging features for glioblastoma survival prediction. Our main goal was to introduce imaging features that are interpretable and that can be replicated across different MR contrasts, scanning equipment or preprocessing. The proposed Hd95 features can be interpreted as measuring the deviation from normal brain morphology due to glioblastoma, and are computed by comparing an automatic whole-brain segmentation with its expected equivalent in healthy subjects. To achieve robustness to missing MR modalities and variations in scanners or acquisition protocols, the automatic segmentations were obtained with a method that was designed to have these properties. 

Using experiments on two different datasets -- one post-operative and one pre-operative -- we showed that the proposed features carry prognostic information and can improve survival models that use conventional clinical features such as age, MGMT and performance status. Group analysis based on the output of our models showed that they could clearly stratify the datasets into low- and high-risk groups with significantly different survival characteristics. Furthermore, individual feature predictiveness was explored, indicating that for some brain structures, very high deformation is a reliable indicator of short survival.

Through feature selection, we discovered several brain structures whose Hd95 value correlates with survival and were therefore retained for training our prediction models. Although the same set of structures was not selected in each case (OS vs. PFS and pre- vs. post-operative), two structures were selected in all three cases: the left pallidum and left putamen. Interestingly, we found that right-sided structures were overall less associated with survival. Two recent studies have explored the association of OS with left-sided glioblastoma (having higher volume in the left hemisphere than the right side) but with contrary results \cite{Abou2019,Yersal2017}. One study \cite{Abou2019}, who showed the association of left-sidedness and worse prognosis, proposed a possible explanation could be that the left hemisphere's functions may be more essential for survival.

As demonstrated in our experiments, the features proposed in this paper readily generalize across datasets: They are independent of scanner and imaging parameters, and they can be computed from both pre- or post-operative images; from data that is skull-stripped or not; and from subjects with missing modalities. We did, however, see worse prognostic performance of the Hd95 features for OS prediction on the pre-operative cohort (BraTS20) compared to the post-operative one (Copenhagen). One possible reason for this discrepancy may be that the BraTS20 dataset contains both glioblastoma and anaplastic astrocytomas, which have different survival characteristics. The fact that BraTS20 is pre-operative may play an important role as well, as the effects of surgery can not be taken into account. 

The proposed Hd95 features measure how much each brain structure is deformed compared to its expected shape in the absence of pathology, and therefore they contain information about the location of the tumor, which has been shown previously to be a prognostic factor for OS \cite{Awad2017, Gorlia2012, Chaichana2008, Gorlia2008, Abou2019, Yersal2017}. Nevertheless, our results show that the proposed Hd95 features carry richer prognostic information for predicting OS than tumor location alone. For predicting PFS, tumor location was found to be a stronger predictor than Hd95 when considering these feature sets individually; however, substantially higher model performance was achieved with a combination of the two, together with clinical and size features. To the best of our knowledge, considerations of tumor location has not been a parameter used in the stratification of patients to treatment in clinical glioblastoma trials, although the poor prognostic feature is recognized in clinical management. Based on our results, the application of survival models exploiting advanced imaging features, such as the ones proposed here, could potentially help minimize bias in stratification in future clinical trials. High quality prognostic information could also potentially guide clinicians in adjusting the intensity of interventions, based on expected outcome and quality-of-life considerations.

While radiomics studies focus on patterns within the tumor region, in this study we have focused on the rest of the brain and ignored the tumor region itself entirely. Using such an approach, we demonstrated that considering out-of-region deformation features together with conventional clinical prognostic factors significantly improves survival models. A recent study \cite{BaeR2018} showed how 18 radiomic features could similarly improve RSF model accuracy when combined with clinical features. Future work may therefore explore combining both within-tumor radiomic features and our Hd95 features to further improve model accuracy.

\bibliography{refs}

\section*{Author contributions statement}

S.P., K.V.L. and I.L. conceived and planned the experiments; S.P carried out the experiments; S.P took the lead in writing the manuscript, closely collaborating with I.L and K.V.L, who supervised the project; S.C. contributed to the design and implementation of the methods; H.P. and T.U. carried out data acquisition and preparation; All authors reviewed the manuscript and provided feedback.

\section*{Funding}
This project has received funding from the European Union’s Horizon 2020 research and innovation program under the Marie Sklodowska-Curie grant agreement No. 765148, as well as from the National Institute Of Neurological Disorders and Stroke under project number R01NS112161.

\section*{Additional Information}
\subsection*{Competing Interests}
The authors declare no competing interests.

\section*{Appendix A}

\begin{figure}[H]
    \centering
    \includegraphics[width=0.81\textwidth]{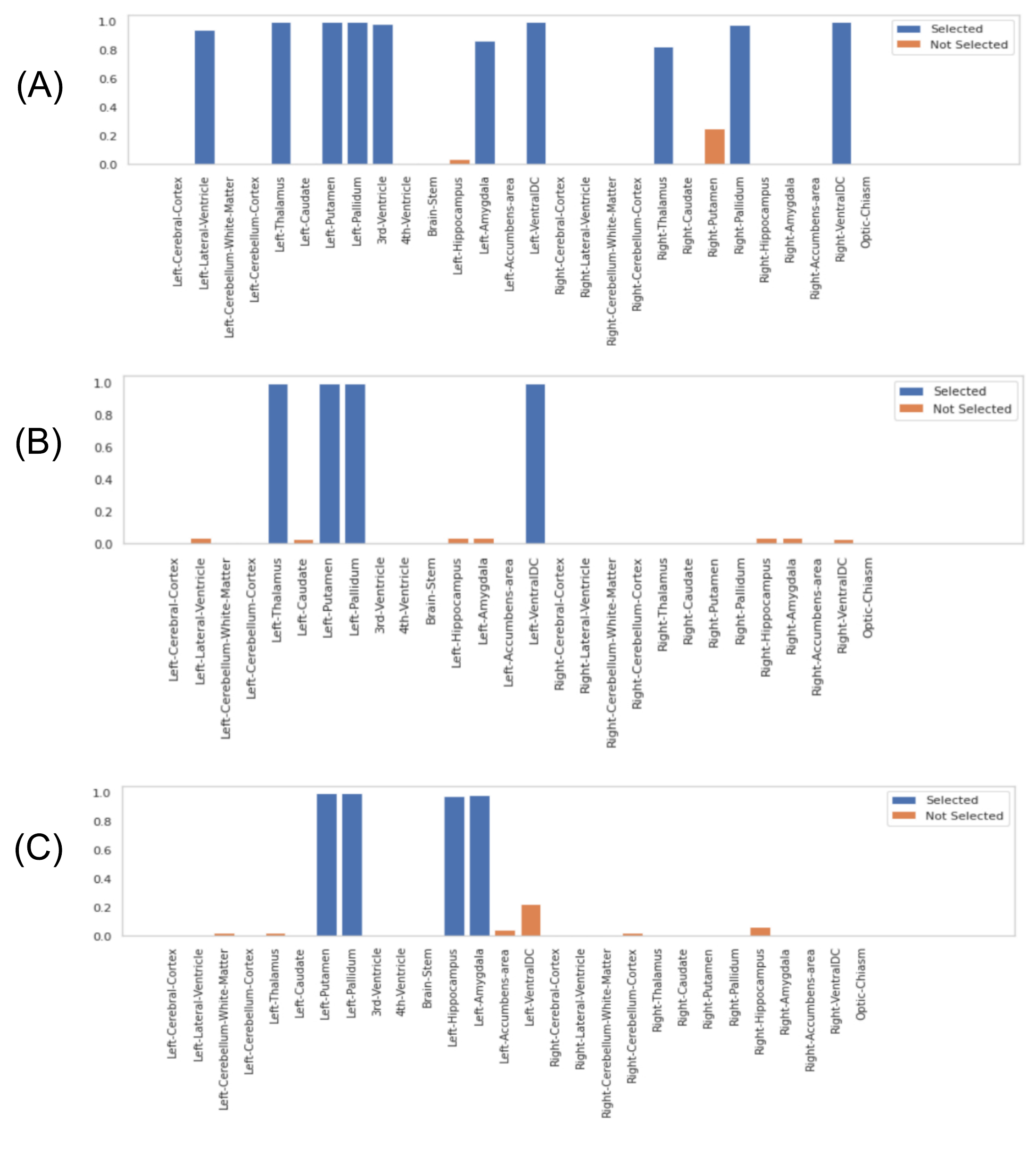}
    \caption{Frequency with which Hd95 features were selected across cross-validation folds on: (A) the Copenhagen data (OS), (B) the Copenhagen data (PFS), and (C) the BraTS20 data (OS). The colors indicate whether the features were also selected when
    a global
    feature selection was performed on the entire dataset instead.}
    \label{fig:selected_os}
\end{figure}

In the interest of conciseness, selection of the proposed Hd95 features was performed on the entire cohort in our experiments, i.e., outside of the cross-validation set-up. While this potentially introduces information leakage between the training and test data within each fold, here we show that the results are only minimally affected in practice. Specifically, we ran our experiments again, selecting the features \emph{within} each fold this time, and recording the number of folds each feature was selected in. Fig.~\ref{fig:selected_os} shows the frequencies (proportion of the cross-validation folds) of selected features -- also shown is a color indicating whether the features were selected on the entire dataset or not. As can be seen from these results, the feature selection is largely consistent across folds, and in alignment with the feature selection performed on the entire cohort.

\end{document}